\documentclass[11pt]{article}
\usepackage{graphics,epsfig}
\usepackage{amsmath,amssymb,amsthm}
\usepackage{url}

\newcommand{\bibnodot}[1]{}

\begin{document}

\title{A Quantum-like Approach to the Stock Market}

\author{Diederik Aerts, Bart D'Hooghe and Sandro Sozzo\\
        \normalsize\itshape
        Center Leo Apostel for Interdisciplinary Studies and Department of Mathematics \\
        \normalsize\itshape
        Brussels Free University, Pleinlaan 2, 1050 Brussels, 
       Belgium \\
        \normalsize
        E-Mails: \textsf{diraerts@vub.ac.be, bdhooghe@vub.ac.be} \\\textsf{ssozzo@vub.ac.be}}
\date{}
\maketitle

\begin{abstract}
\noindent Modern approaches to stock pricing in quantitative finance are typically founded on the \emph{Black-Scholes model} and the underlying \emph{random walk hypothesis}. Empirical data indicate that this hypothesis works well in stable situations but, in abrupt transitions such as during an economical crisis, the random walk model fails and alternative descriptions are needed. For this reason, several proposals have been recently forwarded which are based on the formalism of quantum mechanics. In this paper we apply the \emph{SCoP formalism}, elaborated to provide an operational foundation of quantum mechanics, to the stock market. We argue that a stock market is an intrinsically contextual system where agents' decisions globally influence the market system and stocks prices, determining a nonclassical behavior. More specifically, we maintain that a given stock does not generally have a definite value, e.g., a price, but its value is actualized as a consequence of the contextual interactions in the trading process. This contextual influence is responsible of the non-Kolmogorovian quantum-like behavior of the market at a statistical level. Then, we propose a \emph{sphere model} within our \emph{hidden measurement formalism} that describes a buying/selling process of a stock and shows that it is intuitively reasonable to assume that the stock has not a definite price until it is traded. This result is relevant in our opinion since it provides a theoretical support to the use of quantum models in finance.
\end{abstract}


\section{Black-Scholes model and random walk hypothesis}
Quantitative derivatives pricing was started by Louis Bachelier who first applied the concepts of Brownian motion to the stochastic processes of finance \cite{bachelier1900}. In this section we briefly resume the basic notions, definitions and results in quantitative finance that are needed for our purposes. Our presentation will be focused on the mostly used financial model, i.e. the Black-Scholes model, on its main hypothesis, i.e. the random walk hypothesis, and on their shortcomings. Thus, we do not pretend to be either complete or exhaustive here.

An \emph{option} is a derivative financial instrument specifying a contract between two parties for a future transaction of an asset at a specific price (the \emph{strike price}). The buyer of the option gains the right, which does not entail the obligation, to engage in this future transaction, whereas the seller incurs the corresponding obligation to fulfill this (possible) transaction at specified price. The price of an option derives from the difference between the \emph{strike price} and the value of the \emph{underlying asset} (which can be a stock, a bond, a currency, etc.) plus a premium based on the time remaining until the expiration of the option. More specifically, if we denote by $V(t)$, $I(t)$ and $Z(t)$ the \emph{total value} of an option, its \emph{intrinsic value} and the \emph{time value}, i.e. the premium for potential to increase in value before expiring, respectively, we have
\begin{equation}
V(t)=I(t)+Z(t),
\end{equation}
where the intrinsic value of the option $I(t)$ is expressed in terms of the current price $S(t)$ of the underlying asset and the strike price $K$ by the formula
\begin{equation}
I(t)=S(t)-K .
\end{equation}

An option which conveys the right to buy (sell) something at a specific price is called a \emph{call} (\emph{put}). Most options have an \emph{expiration date}. If the option is not exercised by the expiration date it becomes void and worthless. An option is called \emph{European} if it can be exercised on expiration, an option is called \emph{American} if it may be exercised on any trading day on or before expiration. The intrinsic value of a call option is $I_{C}=\max \left\{ S-K,0\right\}$, while the intrinsic value of a put option is $I_{P}=\max \left\{K-S,0\right\}$. If the intrinsic value of an option is positive, the option is called \emph{in-the-money}, otherwise it is called \emph{out-of-the-money}. 

The theoretical value of an option is evaluated according to specific mathematical models. They aim to predict how the value of an option changes in response to changing conditions. For example, how the price changes with respect to changes in time to expiration or how an increase in volatility would have an impact on the value. Roughly speaking, option pricing theory studies the problem of the `fair' price of option at time $t$ before expiration date $\bar{t}$.

The most widely used model of a financial market for the price of European put and call options is the \emph{Black-Scholes model} \cite{blackscholes1973}, which rules the price of the option over time. The basic idea behind this model is constructing a perfect hedged position, whose value would not depend on the price of the underlying asset (typically, a stock) in order to eliminate risk. 

The Black--Scholes model introduces explicitly the following assumptions:

(i) there is no arbitrage opportunity, i.e. there is no way to make a riskless profit;

(ii) there is a known constant risk-free interest rate at which money can be borrowed and lent;

(iii) stocks can be traded costfree and in any amount;

(iv) the underlying asset does not pay a dividend;

(v) the stock price follows a geometric Brownian motion with constant drift and constant volatility.

Assumption (v) implies that the price $S$ of the underlying asset is a stochastic process satisfying the following differential equation:
\begin{equation} \label{randomwalk}
dS=\mu S dt+\sigma SdW,
\end{equation}
where $W$ is a Wiener process of Brownian motion (a good discrete analogue for $W$ is a simple \emph{random walk}), and $\mu$ (denoting the \emph{drift}) and $\sigma $ (denoting the \emph{volatility}) are constants. Now, if one introduces the \emph{risk-free rate of return} $r$, then one can derive the \emph{Black-Scholes equation} that must be satisfied by the price $V=V(S,t)$ of the option:
\begin{equation} \label{blackscholesequation}
\frac{\partial V}{\partial t}+\frac{1}{2}\sigma ^{2}S^{2}\frac{\partial ^{2}V}{\partial S^{2}}+rS\frac{\partial V}{\partial S}-r V=0.
\end{equation}
As an example, we report the Black-Scholes formula\footnote{Successively, Merton expanded the option pricing model by Black and Scholes and coined the term \emph{Black-Scholes options pricing model}. In 1997 Merton and Scholes were awarded the Nobel Prize in Economics (because of his death in 1995, Black was inelegible for the prize).} concerning the price of a European call option for a non-dividend paying underlying stock. It is given by:
\begin{equation} \label{call}
C\left( S,t\right)  = N\left( d_{1}\right) S-N\left( d_{2}\right) Ke^{-r\left( T-t\right) }.
\end{equation}
The price of a corresponding put option is instead given by:
\begin{equation} \label{put}
P\left( S,t\right)=Ke^{-r\left( T-t\right) }-S+C\left( S,t\right)=N\left( -d_{2}\right) Ke^{-r\left( T-t\right) }-N\left( -d_{1}\right) S.
\end{equation}
The parameters $d_1$ and $d_2$ in Eqs. (\ref{call}) and (\ref{put}) are defined as
\begin{equation}
d_{1} =\frac{\ln \left( \frac{S}{K}\right) +\left( r+\frac{\sigma ^{2}}{2} \right) \left( T-t\right) }{\sigma \sqrt{T-t}}
\end{equation}
and
\begin{equation}
d_{2} =\frac{\ln \left( \frac{S}{K}\right) +\left( r-\frac{\sigma ^{2}}{2} \right) \left( T-t\right) }{\sigma \sqrt{T-t}} = d_{1}-\sigma \sqrt{T-t} ,
\end{equation}
respectively, while $N\left( t\right)$ is the cumulative distribution function of the standard normal distribution 
\begin{equation}
N(t) =\frac{1}{\sqrt{2\pi }}\int_{-\infty}^{t} e^{-z^{2}/2} dz \ .
\end{equation}

It should be observed that the model presented above contains several simplifications and idealizations that can hardly be realized in the real stock market. Moreover, the claimed independence of the Black-Scholes model from risk is only theoretical, since following the model exposes the user to some unexpected risk. Notwithstanding this, empirical tests generally show a good accordance with the predictions following from the model, so that the Black-Scholes model is, together with its refinements to real cases,\footnote{A variant of the Black-choles model, called the \emph{binomial option pricing model}, was presented in 1979 \cite{coxrossrubinstein1979}. This model assumes that price movements follow a binomial distribution and provides a discrete time approximation to the continuous process underlying the Black-Scholes model. In the case of European options without dividends, the binomial model value converges on the Black-Scholes formula value as the number of time steps increases. Although computationally slower than the Black-Scholes formula, the binomial model is more accurate, particularly for longer-dated options on securities with dividend payments.  For these reasons, various versions of the binomial model are widely used in option markets.} the predominant model of option pricing in quantitative finance.

A crucial assumption for the derivation of Eq. (\ref{blackscholesequation}) is the \emph{random walk hypothesis} underlying Eq. (\ref{randomwalk}), that is, the assumption that assets prices evolve according to a classical random walk. Roughly speaking, the closing price of an asset in a financial market, such as Wall Street, should be determined by a coin flip, hence it should be a classical Kolmogorovian process. This random walk hypothesis has characterized the financial derivatives modelling and valuation in the banks and financial institutions, and is strictly linked with the so-called \emph{efficient market hypothesis}, that is, the assumption that financial markets are `informationally efficient' and prices of traded assets instantly change to reflect new public information. It is evident that both random walk and efficient market hypotheses refer to ideal situations, but their validity rests on the fact that it is practically impossible to exploit market inefficiencies at the level of assets prices.

However, in the last years several authors have pointed out that there are some situations in finance where the deviations from the Black-Scholes predictions cannot be attributed to random fluctuations but, rather, some of the underlying hypotheses of the model could not hold \cite{choustova2001,haven2002,schaden2002,haven2003a,haven2003b,baaquie2004,bagarello2006,bagarello2007,choustova2007,yehuang2008,ataullahdavidsontippett2009}. In particular, the random walk hypothesis is put at stake. These authors maintain, by referring to existing experimental data, that the random walk model works well in static situations where no phase transition occurs. Whenever something happens, such as in a \emph{stock market crash}, the random walk model fails, while coherent effects appear. In these exceptional situations, deviations are not averaged out, but seem to reinforce themselves, and people start acting as a group with correlated  actions (\emph{herding}). As a consequence, alternative models to the Black-Scholes equations have been proposed and some descriptions have been inspired by quantum mechanics \cite{choustova2001,haven2002,schaden2002,haven2003a,haven2003b,baaquie2004,bagarello2006,bagarello2007,choustova2007,yehuang2008,ataullahdavidsontippett2009}. An analogy is assumed with statistical mechanics. It is indeed well known that if we consider a gas with many particles and the temperature is high, then the gas is nondegenerate and the classical Maxwell-Boltzmann distribution is a good approximation of its behavior. Viceversa, if the temperature becomes very low, then quantum coherence and other quantum effects appear, the gas becomes degenerate and must be described within quantum, e.g., Fermi-Dirac or Bose-Einstein, statistics. These scholars suggest that similar phenomena occur in the stock market (herding, stampede, etc.) whenever some parameters reach critical values. We will prove in the next sections that a similar conclusion can be attained if one adopts a different perspective, and explain why a quantum-like formalism is needed to model financial systems.

\section{Potentiality structures in finance}
We have seen in the previous section that standard option pricing models are based on classical stochastic processes. In particular, the value of an option is a suitable function of some set of random variables. One could then observe that the formalism of quantum mechanics and the quantum probability model could supply more general and flexible tools to model the pricing of financial assets (options, stocks, shares, etc.). But, there are stronger reasons for using quantum structures in finance.

It is a basic doctrine in quantum mechanics that a property, e.g., the value of an observable, of a physical entity in a given state cannot be considered as a preexisting feature of the entity, as it is usually assumed in classical physics. On the contrary, the property is actualized in a measurement process and can be different if a different measurement context is considered. Therefore, the measurement context introduces a nondeterministic change of state of the entity that is measured as a consequence of the interaction between the entity itself and the measuring apparatus. Moreover, in a measurement a transition occurs from `potential' to `actual', that is, a property that was only potential before the measurement becomes actual after it, and viceversa. One typically says in this case that the measured entity is in a \emph{potentiality state} with respect to the measurement context. As a consequence, quantum probability cannot be interpreted as formalizing the subjective ignorance about the actual state of the entity, as it occurs for classical (Kolmogorovian) probability and statistical mechanics, but it is rather caused by the presence of a lack of knowledge (\emph{fluctuations}) concerning how context interacts with the entity that is considered. The ensuing formalization of this \emph{quantum potentiality} entails that the quantum probability model is non-Kolmogorovian \cite{accardi1982,accardifedullo1982,pitowsky1989}. The Brussels group has inquired into the conceptual foundations of quantum mechanics for years to provide an operational axiomatization of this theory. This research led the authors to elaborate a \emph{State Context Property} (\emph{SCoP}) formalism to describe any (not necessarily physical) entity \cite{aerts1999a,aerts1999b,aerts2002b}. The SCoP formalism recovers classical and quantum mechanics generalizing both of them, and has been used in previous investigations to model concepts and their combinations \cite{aertsgabora2005a,aertsgabora2005b}. The probability arises in it as a consequence of the lack of knowledge about how context influences the entity under study (\emph{hidden measurement formalism}) \cite{aerts1986,aerts1993,aerts1995,aerts1998,aerts2002a}. This formalism is able to describe any kind of quantum-like potentiality structures.

An example of the above situations has been found outside physics, i.e. the opinion poll model \cite{aertsaerts1994}. A quantum-like potentiality structure occurs in an opinion poll because two types of questions are possible.

{\it Type I}: e.g., the question `Are you a smoker?': answer `yes' or `no' is predetermined, independent of the happening of the opinion poll. 

In this case, the answer only changes the knowledge of the interrogator, hence it is a question with a predetermined answer ({\it classical question}).

{\it Type II}: a question where the answer is not necessarily predetermined, as follows.

(i) `Are you a proponent of the use of nuclear energy?'.

(ii) `Do you think it would be a good idea to legalize soft-drugs?'.

(iii) `Do you think it is better for people to live in a capitalistic system?'.

In these cases, the opinion of an interviewee is possibly not determined before the opinion poll takes place, but it is formed during the process of the opinion poll. Hence, the interviewee's mind starts from a potentiality state, and the overall interaction process during the `measurement of the opinion' realizes a final answer `yes' or `no'. We say that this  type of question is {\it nonclassical}. It has been shown that the ensuing probability structure is non-Kolmogorovian and admits a generalized quantum or, equivalently, quantum-like, representation \cite{aertsaerts1994}.

A careful analysis shows that similar situations can be found in the stock market. Indeed, it is reasonable to suppose that prices of assets are not predetermined, but they depend on the simultaneous happenings of human decisions to buy and/or sell. In this perspective, the structural and dynamical nature of an asset and its relation with `the price of the asset' is similar to nonclassical questions in the quantum opinion poll model. More specifically, the equivalent of a classical question with a predetermined answer is a question related to the price of a common material object sold in a shop, where the price is fixed apriori. As a consequence, a classical model is satisfactory for the buying and selling of normal goods, for one likes thinking that the prices of goods are also given when no selling/buying takes place.\footnote{We note that there may be situations where also the buying and selling of normal goods follow a nonclassical pattern. Imagine, e.g., the following situation. In a shop there is still one item for sale, with a price written on the item. A person enters the shop, and says to the owner that he or she wants to buy this item. Then, a second person enters the shop and says that he or she wants to buy this item. But the shop owner replies that the item is already sold. Now, the second person could state that he or she wants to offer more money. Then different things might happen. Perhaps the owner does not like at all that bargaining about a price takes place in his or her shop, and thus says that the item is sold, and there the story stops. This would happen in case the owner considers his or her shop really a shop with items that have their price also when no selling/buying is taking place. And hence, without being aware of this, such an owner in fact introduces a classical structure with respect to price in his or her shop. Because indeed, in a typical popular market place in open air prices are not considered to be independent of the selling/buying moment, hence nonclassical situations naturally occur. This example of how buying and selling and their relation with respect to prices are in an open air popular market shows that the general state of affairs is quantum-like, and the classicality with respect to `serious shops' introduces an extra element, not intrinsically connected to the buying/selling relation with price.} On the contrary, the equivalent of a nonclassical question with a non-predetermined answer is a question related to the price of an asset in the financial market, where there is a `bidding for the price' during the trading process. Notwithstanding this, in standard economic theories one usually refers to the efficient market hypothesis (see the first section). If information is spread `efficiently', or locally, prices are set in a correct way, and the market evolves toward a state where the efficient market hypothesis is again satisfied. In this case, which is typical of a normal market situation, a nonclassical effect due to the non-predetermined price of an asset and contextual influence by decision context `wanting to buy and/or sell it' is small. Thus, the use of classical Black-Scholes equations for the assets prices seems to be justified.

The situation is deeply different if one considers what happens during a market crash or in an economical crisis. In these cases, indeed, information about buying/selling is spread non-locally, the efficient market hypothesis does not hold any longer, and the nonclassical effects due to contextual decision making cannot be neglected. As a consequence, one could say that classical stochastic equations only describe the average evolution of option prices if `nothing special happens'. On the contrary, if `catastrophic' events (crash), or situations of high volatility, occur, the nonclassical effects appear, and quantum structures are necessary.

The deep reason why the price of an asset is not a priori fixed but it is realized in a trading process will be clear from the following.

 \section{A SCoP formalism to model the stock market} 
The analysis undertaken in the previous section suggests that a more refined model of the stock market could be based on the quantum formalism, or on a generalization of it. In this section the SCoP formalism will be presented to put forward such a possible generalization.

Suppose that the entity $T$ under investigation is a stock. At a specific moment, the stock $T$ is described by a pure state $p$ representing the reality of $T$ at that time. A context $e$ for the stock $T$ is a part of the outside reality influencing $T$ in such a way that the state $T$ is changed. The context $e$ describes, for example, the whole of humans' decisions in a trading process, in which $T$ is bought/sold. In general, the interaction between the stock $T$ and the context $e$ determines a change of the state $p$ of $T$ to a state $q$ (by analogy with the state transition of a quantum entity in a measurement process, we call \emph{collapse} the state transition of $T$). We denote the set of states and contexts for $T$ by $\Sigma$ and ${\mathcal M}$, respectively. If $p$ does not change under the context $e$ we say that $p$ is an \emph{eigenstate} for $e$, otherwise we say that $p$ is a \emph{potentiality state} for $e$. The probability that the state $p$ changes to state $q$ under the context $e$ is denoted by $\mu(p,e,q)$. Obviously, a state transition of this kind is generally a stochastic process.

A property $a$ of a stock $T$ is an element of reality that $T$ can possess independently of the context influencing $S$. A property $a$ for the stock $T$ is, for example, the price of the stock at a given time. We denote the set of properties of $T$ by ${\mathcal L}$. If the stock $T$ possesses a property $a$ we say that $a$ is \emph{actual} for $T$, otherwise we say that $a$ is \emph{potential}. A stock $T$ in a given state $p$ possesses a set of properties that are actual. Properties that are actual for $T$ being in a specific state can be potential when $T$ is in another state. This defines a function $\xi: \Sigma \longrightarrow {\mathcal P}({\mathcal L})$, where ${\mathcal P}({\mathcal L})$ is the set of all subsets of $\mathcal L$, which maps each state $p$ into a set $\xi(p)$ of all properties that are actual in this state. By introducing the function $\xi$ we replace the locution ``the property $a$ is actual when the stock $T$ is in the state $p$'', by ``$a \in \xi(p)$''. If the state $p$ of $T$ is changed under influence of the context $e$ into the state $q$, then the set $\xi(p)$ of actual properties in state $p$ is changed into the set $\xi(q)$ of actual properties in state $q$. In particular, we say that the stock $T$ in the state $p$ \emph{displays} a price ranging between the value $x$ and the value $y$ in a trading process if the property $a$: `having a price between $x$ and $y$' is actual when $T$ is in the state $p$. Then, the trading process changes the state $p$ of $T$ into a state $q$ which is such that $a\in \xi(q)$.

Having introduced the foregoing basic definitions and mappings, we are now ready to associate a given \emph{stock} $T$ with a \emph{SCoP system}, that is, a set $\Sigma$ of states, a set $\mathcal M$ of contexts, a set $\mathcal L$ of properties, and two functions $\mu$ and $\xi$. The function $\mu$ is defined as $\mu: \Sigma \times {\mathcal M} \times \Sigma \times {\mathcal M} \longrightarrow [0,1]$, $(q,f,p,e) \mapsto \mu(q,f,p,e)$, where $\mu(q,f,p,e)$ is the probability that the stock in state $p$ changes to state $q$ under influence of the context $e$, and such that the context $e$ changes to context $f$. Hence $\mu$ describes the structure of the contextual interaction of the stock under study. The function $\xi$ is defined as $\xi: \Sigma \longrightarrow {\mathcal P}({\mathcal L})$, $p \mapsto \xi(p)$, where $\xi(p)$ denotes the set of all properties that are actual in state $p$. Hence $\xi$ describes the internal structure of the entity or, better, how properties depend on the different states the stock can be in. 
The above definition can be extended to different interacting stocks in order to describe a financial market.

It follows from the preceding treatment that a given stock has generally not a definite price, even if it is in a definite state at a specific instant of time. On the contrary, the stock's price is realized in a trading process, which is generally a stochastic process and depends on the contextual interactions in humans' decisions of buying/selling. The analogy with the behavior of a quantum entity interacting with a measurement context is thus evident, and supports the idea that a quantum-like formalism should be employed to model stocks transformations.

\section{A sphere model for a stock\label{spheremodel}}
The nonclassical character of stock prices can be better understood by elaborating a macroscopic contextual model that reproduces the fluctuations of a given stock in a financial market. We will adapt a model that has already been worked out and used to identify the presence of quantum structures, and has been called the {\it sphere model} \cite{aerts1993,aerts1995,aerts1998}.

The sphere model consists of a physical entity $T$ that is a material point particle $P$ moving on the surface of a sphere that we denote by $surf$, with center $O$ and radius 1. The entity $T$ describes the stock that is considered. The unit vector ${v}$ where $P$ is located on $surf$ represents the pure state $p_v$ of $T$ (see Fig. 1a). For every point ${u} \in surf$, the following measurement $e_u$ can be introduced. Let $-{u} \in surf$ be the diametrically opposite point with respect to $u$, and let us install a piece of elastic of 2 units of length such that it is fixed with one of its endpoints in ${u}$ and the other endpoint in $-{u}$. Once the elastic is installed, the particle $P$ falls from its original place ${v}$ orthogonally onto the elastic, and sticks on it (Fig. 1b). Then the elastic breaks somewhere and $P$, attached to one of the two pieces of the elastic (Fig. 1c), moves to one of the two endpoints ${u}$ or $-{u}$ (Fig. 1d). Depending on whether the particle $P$ arrives at ${u}$ (as in Fig. 1) or at $-{u}$, we attribute the outcome $o_{1}^{u}$ or $o_{2}^{u}$ to $e_{u}$. The elastic installed between  ${u}$ and $-{u}$ plays the role of a (measurement) context for the entity $T$. The measurement $e_u$ describes a process of trading of the given stock. More precisely, we say that the outcome $o_{1}^{u}$ is obtained if the price of the stock ranges between the values $x$ and $y$. Conversely, we say that the outcome $o_{2}^{u}$ is obtained if the price of the stock lies outside the interval between $x$ and $y$.   
\begin{figure}[h]
\begin{center}
\includegraphics[scale=0.75]{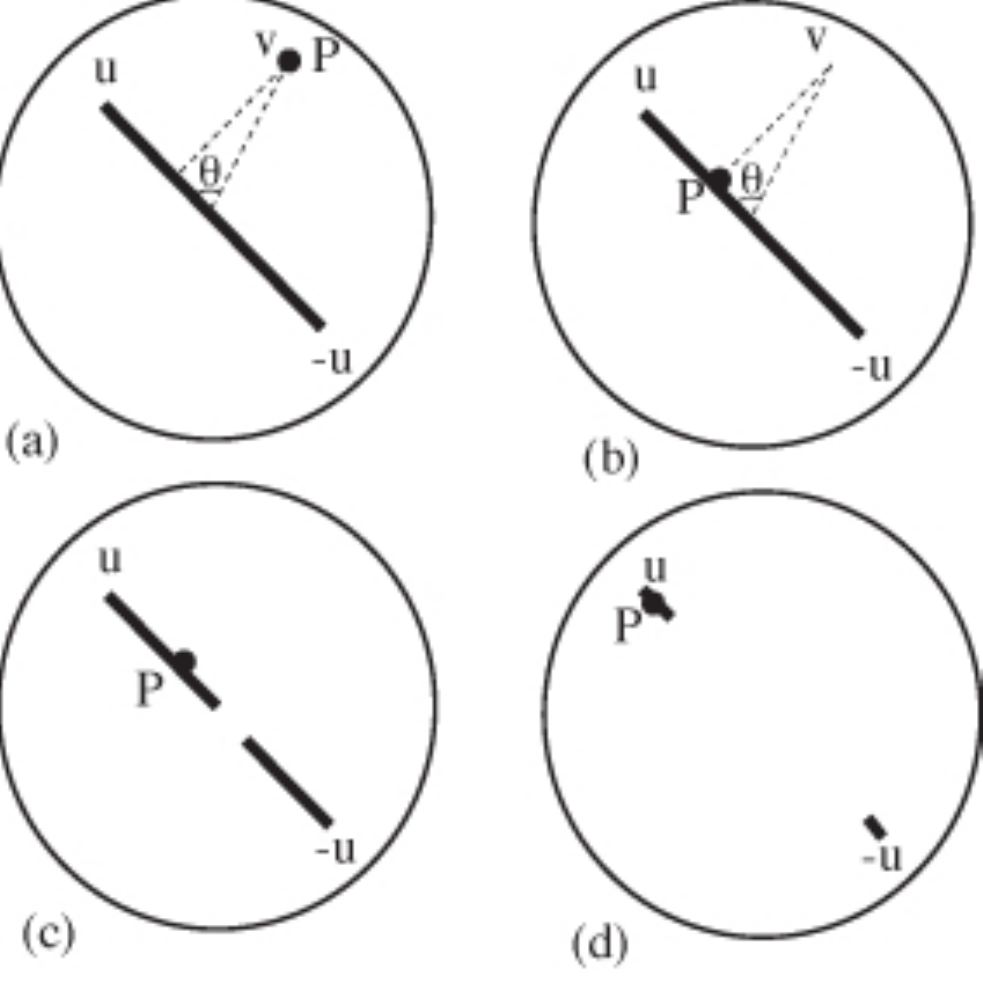}
\caption{A representation of the sphere model}
\end{center}
\end{figure}
Let us now consider elastics that break in different ways depending on their physical construction or on other environmental happenings. We can describe such a general situation by a classical probability distribution
\begin{equation}
\rho: [-u,u] \longrightarrow [0, +\infty]
\end{equation}
such that 
\begin{equation}
\int_{\Omega} \rho(x) d x  
\end{equation}
is the probability that the elastic breaks in the region $\Omega \subset [-u, u]$. We also have:
\begin{equation}
\int_{[-u,u]} \rho(x) d x =1, 
\end{equation}
which expresses the fact that the elastic always breaks during a measurement. A measurement $e_{u}$ characterized by a given distribution $\rho$ will be called a  \emph{$\rho$-measurement} and denoted by $e_{\rho}^{u}$ in the following. A $\rho$-measurement is a hidden measurement for the entity $T$, in the sense that the probabilities associated with the outcomes $o_{1}^{u}$ and $o_{2}^{u}$ can be interpreted as due to $\rho$. Then, the distribution $\rho$ describes the lack of knowledge about humans' decisions with respect to the given trading process, hence the subjective lack of knowledge about how humans interact with respect to the given stock.

Let us calculate the probabilities involved in a $\rho$-measurement. The transition probabilities that the particle $P$ arrives at point ${u}$ (hence the outcome of the measurement is  $o_{1}^{u}$) and $-{u}$ (hence the outcome of the measurement is $o_{2}^{u}$) under the influence of the measurement $e_{\rho}^{u}$, are respectively given by:
\begin{equation} \label{transition}
\mu_{\rho}(p_{u}, e_{u}, p_{v})=\int_{-1}^{{v} \cdot {u}} \rho(x) d x \quad {\rm and} \quad
\mu_{\rho}(p_{-u}, e_{u}, p_{v})=\int_{{v} \cdot {u}}^{1} \rho(x) d x .
\end{equation}

Therefore, the process of buying/selling of the stock $T$ can be considered as the result of a contextual interaction between $T$ itself and the surrounding conceptual landscape made up of the agents' decisions. This sentence needs some comments. The fact that the decisions taken by humans with respect to specific situations have a nonclassical character is manifestly revealed by several paradoxes in economics and decision theory (\emph{Allais} and \emph{Ellsberg paradoxes}, \emph{conjunction fallacy}, \emph{disjunction effect}, etc.). These drawbacks show that real subjects are not only guided by classical logic and probability theory in their decisions. On the contrary, subjects also follow a different way of thinking, which has a conceptual nature, depends on their contextual interaction with the given situation, and is structurally connected with the quantum formalism, as we have argued in a number of papers \cite{aerts2009,aertsdhooghe2009,aertssozzo2011,aertsdhooghesozzo2011}. More specifically, in these investigations, we have put forward the hypothesis of the presence of two structured and superposed layers in human thought: a \emph{classical logical layer}, that can be modeled by using a classical Kolmogorovian probability framework, and a \emph{quantum conceptual layer}, that can instead be modeled by using the probabilistic formalism of quantum mechanics. The thought process in the latter layer is given form under the influence of the totality of the surrounding conceptual landscape, hence context effects are fundamental in this layer \cite{aerts2009,aertsdhooghe2009}.

It is important to observe that the probabilities in Eq. (\ref{transition}) cannot be cast into a unique Kolmogorovian scheme. This result can be proved by referring to \emph{Pitowsky's polytopes}, or to \emph{Bell-like inequalities} \cite{pitowsky1989,aerts1986}. Furthermore, if we limit ourselves to consider uniform probability distributions $\rho$, then the probabilities in Eq. (\ref{transition}) become the standard quantum probabilities for spin measurements, since our sphere model is a model for a spin 1/2 quantum particle \cite{aerts1993,aerts1995,aerts1998}. A proof that the probability model of a spin 1/2 quantum particle is not Kolmogorovian is due to Accardi and Fedullo \cite{accardifedullo1982}, and we refer to their paper for technical details.

The sphere model presented above is, obviously, idealized and oversimplified. But it already shows that the monetary transitions in a trading process for a stock cannot be modeled by using classical Kolmogorovian probabilities, because of their intrinsic contextuality, and that a non-Kolmogorovian quantum-like framework is necessary. Moreover, the presence of elastics is, in our opinion, a good illustration to express the intuition that, when an asset is not bought/sold, there is a kind of `tension on the price of the asset'. This tension reproduces what happens in the moments between the times in which a buying/selling takes place. The breaking of the elastics then describes the price assignment to the asset, that is, the collapse.

\section{Conclusions}
Modern quantitative finance rests on classical stochastic models of market behavior. More precisely, the Black-Scholes equations are used to model prices of assets, options and stocks, their solutions rule the stock market at equilibrium, and deviations from these solutions are usually considered as random fluctuations from equilibrium. The Black-Scholes equations are founded on the random walk hypothesis. But an explanation does not exist either for the utilization of this hypothesis or for its violation in specific situations. It can be shown by referring to the experimental data existing in the literature that the random walk model is a good approximation of the market reality in static situations, that is, when no special, i.e. out of the ordinary set of events with respect to stock market dynamics, takes place. Whenever such extra-ordinary events take place, e.g., events leading to financial and/or economical crisis, the random walk model fails. Often coherent effects start to appear. Eventually it occurs that, for example, a herd dynamics is established, or a destructive stampede comes into being. We argue that a situation similar to quantum statistics may occur in finance: whenever news of world importance is spread around, it cognitively aligns human choices and preferences, which creates a quantum-like coherence, hence nonclassical effects. In this case, the quantum coherence is generated by the global nature of the news items. If news items remain local, in the sense that their effects are not distributed over the world globe, then quantum coherence cannot emerge, and the stock market is well approximated by a classical random walk. On the contrary, whenever news is non-local and has a global effect, e.g., in a financial crisis, coherent quantum-like phenomena may occur, such as herding or stampede.

Basing ourselves on a SCoP formalism, which supplies an operational foundation of quantum mechanics, and on a hidden measurement formalism, which interprets quantum probabilities as due to fluctuations in the interaction between entity and context, we have provided strong arguments to maintain that a given stock has no definite price when it is not traded, whereas its price is actualized whenever a buying/selling process occurs. This actualization is a consequence of the overall conceptual landscape made up of the whole of humans' decisions. In some cases, it may occur that these contextual effects can be neglected, thus classical models can be applied. But, relevant situations also exist in which these contextual effects cannot be neglected. A quantum-like description is instead needed in these cases. The process of trading in a stock market has been concretely illustrated by means of an elastic sphere model which is contextual and has a non-Kolmogorovian quantum-like structure. As a consequence, our approach provides a theoretical support to the development of quantum models for the stock market.

The results presented in this paper are part of an ongoing activity of our Brussels research group to identify the classical or non classical nature of phenomena that are traditionally studies in different domains of science \cite{aertsaerts1994,gaboraaerts2002,aertsczachor2004,aertsgabora2005a,aertsgabora2005b,aerts2009,aertsdhooghe2009,aertssozzo2011,aertsdhooghesozzo2011}. In this respect we also mention that there is actually a successfully growing international research activity identifying quantum structures in domains of science different from the micro-world. Interesting results have been obtained with respect to economics \cite{choustova2001,haven2002,schaden2002,haven2003a,haven2003b,baaquie2004,bagarello2006,bagarello2007,choustova2007,yehuang2008,ataullahdavidsontippett2009} as we mentioned already in this paper. But more extensive results where obtained in psychology and cognitive science, more specifically in the modeling of concepts and their combinations \cite{gaboraaerts2002,aertsgabora2005a,aertsgabora2005b,aerts2009,aertssozzo2011,aerts2007,kittoetal2010,bruzaetal2011,galeaetal2011}, and in decision theory \cite{aertsaerts1994,Busemeyer2006,Khrennikov2008,Accardi2009,Khrennikov2009,Pothos2009,Busemeyer2009,Lambertmogilianski2009,franco2009}, and in situations related to semantic spaces 
\cite{aertsczachor2004} such as problems information retrieval \cite{widdows2003,vanRijsbergen2004,widdows2006,Melucci2011}.


\end{document}